\documentstyle[11pt,aaspp4]{article}

\lefthead{ B.\ D.\ Wandelt }
\righthead{Primordial Non-Gaussianity...}

\begin{document}
\slugcomment{
\begin{tabular}{rr}
Imperial/TP/97-98/19\\
TAC/1998--001
\end{tabular}
}
\title{ Primordial Non-Gaussianity:
Baryon Bias and Gravitational Collapse of Cosmic String Wakes}

\author{ Benjamin D. Wandelt}
\affil{ Theoretical Astrophysics Centre\\
	Juliane Maries Vej 31,
	DK-2100 Copenhagen,
	Denmark\\
	and\\
	Theoretical Physics Group, Blackett Laboratory\\
	Imperial College,
	Prince Consort Road,
	London SW7 2BZ,
	UK\\
}

\newcommand\ba{\begin{array}}
\newcommand\ea{\end{array}}
\newcommand\bc{\begin{center}}
\newcommand\ec{\end{center}}
\newcommand\be{\begin{enumerate}}  
\newcommand\ee{\end{enumerate}}  
\newcommand\bi{\begin{itemize}}  
\newcommand\ei{\end{itemize}}  
\newcommand\bd{\begin{description}}  
\newcommand\ed{\end{description}}  
\newcommand\beq{\begin{equation}}  
\newcommand\eeq{\end{equation}}  
\newcommand\beqa{\begin{eqnarray}}  
\newcommand\eeqa{\end{eqnarray}}  

\newcommand\gap{\smallskip}
\newcommand\eqbox[1]{\fbox{\rule[-.8em]{0em}{2.3em}$\displaystyle\ #1\ $}}

\newcommand{\dum}{\thechapter.}
\newcommand{\eq}[1]{Eq.\ (\ref{#1})}
\newcommand{\eqs}[2]{(\ref{#1}--\ref{#2})}
\newcommand\cf{{c.f.\ }}
\newcommand\eg{{\em e.g.,\ }}
\newcommand\etc{{\em etc}}
\newcommand\ie{{\em i.e.,\ }}
\newcommand\qed{\hfill {\bbox{ QED}}}

\newcommand\A{\op A}
\newcommand\B{\op B}
\newcommand\mathC{\mkern1mu\raise2.2pt\hbox{$\scriptscriptstyle|$}
                {\mkern-7mu\rm C}}
\newcommand\D{\op D}
\renewcommand\H{{\cal H}}                       
\renewcommand\P{\op P}                          
\newcommand\Q{\op Q}
\newcommand\U{\op U}
\newcommand\Z{{\rm \bf Z}}

\newcommand\abs[1]{\vert {#1}\vert}
\newcommand\dby[1]{{d\over d#1}}
\newcommand\down{\vert\!\downarrow\,\rangle}

\newcommand\half{{\frac 12}}
\newcommand\hhalf{{\textstyle\half}}
\newcommand\ioh{{i\over\hbar}}

\newcommand\ket[1]{\,\vert#1\rangle}
\newcommand\lleft{\vert\!\leftarrow\rangle}
\newcommand\map{\longrightarrow}
\newcommand\mathR{{\rm I\! R}}
\newcommand\norm[1]{\parallel\!\v#1\!\parallel}
\newcommand\op[1]{\widehat{#1}}
\newcommand\ra{\rangle}
\newcommand\rright{\vert\!\rightarrow\rangle}
\newcommand\tr{{\rm tr}\,}
\newcommand\triple[3]{\langle#1\vert\,#2\,\vert#3\rangle}
\newcommand\twid[1]{\tilde {#1}}
\newcommand\unit{{\rm I}}
\newcommand\up{\vert\!\uparrow\,\rangle}
\renewcommand\v[1]{{\bf{#1}}}        
\newcommand\avg[1]{{\langle{#1}\rangle}} 
 
\newcommand\<[2]{\langle\v#1,\v#2}              
\renewcommand\>{\rangle}                        
\renewcommand\[{[\,}                            
\renewcommand\]{\,]}                            

\renewcommand\a{\alpha}                         
\renewcommand\b{\beta}                          
\newcommand\g{\gamma}
\newcommand\de{\delta}
\newcommand\e{\varepsilon}
\newcommand\z{\zeta}
\newcommand\k{\kappa}
\renewcommand\l{\lambda}                        
\newcommand\m{\mu}
\newcommand\n{\nu}
\newcommand\r{\rho}
\newcommand\s{\sigma}
\newcommand\th{\theta}
\newcommand\Th{\Theta}
\newcommand\f{\phi}
\newcommand\w{\omega}
\renewcommand\O{\Omega}				
\renewcommand\L{\Lambda}			
\newcommand\bolde{\epsilon\mkern-6mu\epsilon}  
\newcommand\nnabla{\nabla\mkern-14mu\nabla}  
\newcommand{\mue}{{\mbox{\boldmath $\mu$}}}

\newcommand\pdby[1]{{\partial_{ #1}}}

\newcommand\pddby[1]{{\partial_{ #1 #1 }}}

\newcommand\rem[1]{$\vert \underline{\overline{\text{{\bf#1}}}}\vert
$}
\newcommand\dmu{\partial_\mu}

\begin{abstract}

I compute the 3-D non-linear evolution of gas and dark matter fluids in the
neighbourhood of 
cosmic string wakes which are formed at high redshift ($z\simeq 2240$)
for a ``realistic'' scenario of wake formation. 
These wakes are the ones which stand out
most prominently as cosmological sheets and are expected to play a
dominant r\^ole in the cosmic string model
of structure formation. Employing a high-resolution 3-D hydrodynamics
code to evolve these wakes until the present day 
yields results for the baryon bias generated in the inner wake
region. I find that today, wakes would be $1.5 h^{-1}$ Mpc thick and
contain a 70\% excess in the density of baryons over the dark matter
density in  their centre. However, high
density peaks in the wake region do not inherit a baryon
enhancement. I propose a mechanism for this erasure of the baryon
excess in spherically collapsed objects based on the geometry change
around the collapsing region. 
Further, I present heuristic arguments
for the consequences of this work for 
large scale structure in the cosmic string model and conclude that the
peculiarities of wake formation are unlikely to have
significant import on the discrepancy between power spectrum
predictions and observations in this model. 
If one invokes the nucleosynthesis bound
on $\Omega_b$ this could be seen as strengthening the case against
$\Omega_m=1$ or for low Hubble constants.

\end{abstract}

\renewcommand{\thefootnote}{\arabic{footnote}}
\setcounter{footnote}{0}

\section{Introduction}
\label{intro}

Two widely studied paradigms attempting to explain the 
origin of structure are ``inflation'' (\cf  \cite{KandT} for
an introduction and references)  and ``topological 
defects'' (\cite{kibble}; \cite{V&S}) such as cosmic strings or
textures. Finding theoretical predictions which lead to 
observational strategies to distinguish between them
or rule any one of them out has been a major field of activity in
modern cosmology.

The family of inflationary cosmogonies (with various
inflaton potentials,
dark matter contents, open, flat and closed topologies) 
give rise to a wide range of physical predictions.
On the other hand,
the non-linear nature of the 
evolution of defect networks have made 
it difficult to pin down theoretical predictions for a long time.
Also, the inherent presence of non-Gaussianity in such models makes
calculating the full predictions for such theories an enormous task,
as two-point functions contain only a limited amount of information about
the resulting density fields.

Recently it was shown that the cosmic string scenario  of
structure formation suffers from difficulties fitting 
the galaxy power-spectrum, when it is 
normalized to COBE (\cite{PST1}; \cite{PST2}; \cite{ABR1},
1997b)\footnote{See \cite{maeh} for a different, earlier perspective.}. 
However, these calculations describe
fluctuations in the dark matter component, which are then compared
to the galaxy powerspectrum. To make contact
between theory and observation, and to respond to what Albrecht et al.
(1997a) term the $b_{100}$ problem, the differences of
baryon and dark 
matter dynamics  must be studied for active perturbation models.

This is particularly interesting in
the case of cosmic strings as non-Gaussianity leads to rich
physics in the wake of a cosmic string (\cite{Vachaspati}; 
\cite{stebbETAL}; \cite{perivolaro}).
The dark-matter dynamics of
cosmic string wakes and their r\^ole for
the generation of large scale structure has been studied actively 
(\cite{Vachaspati1}; \cite{VachVil}; \cite{A&Sa}, 1992b; \cite{RobAlb})
The non-linear evolution of
gas/CDM--wakes in three dimensions has first been
studied by \cite{Sornborger}. The authors consider two extreme
cases: a fast-moving ($v=0.5 c$) 
straight string, modeled by a planar symmetric, pure velocity
perturbation 
and a slow-moving $(v<<c)$ wiggly string representing a Newtonian
line source potential. 
Their most interesting finding is an enhancement in the
baryon overdensity compared to the dark matter overdensity (ie.\ a
{\em baryon bias}) in the centre of the wake. This excess amounted to 
about 140\% in the
case of a fast moving string. In the case of early
wakes, which are formed around the time of matter radiation equality,
the baryon overdensity is initially more spread out and the opposite
is true: a sheet of excess dark matter is sandwiched between two
``baryonic'' layers. Later, as the gravity of the dark matter begins
to dominate, the baryonic matter clumps in the core of the wake,
reversing to baron enhancement in the core.

This beautiful effect is caused by trapping  the collisionless 
dark matter in an oscillatory mode about the wake centre, while the
baryons shock--heat and build a pressure supported peak in the core of
the wake \footnote{For early wakes, the initial sound speed is high and a weak
shock forms spreading the baryon overdensity beyond the width of the
dark matter.}. The oscillatory mode prevents the full gravitational collapse
of the dark matter as long as the planar symmetry is maintained.

It is worth stressing that 
were their results to generalise to less restrictive conditions, 
a sizeable baryon bias
would be a signal for the presence of
non-Gaussianities in the very early universe. 
In high-resolution hydrodynamics simulations (\cite{virgo}), 
none of the tested
Gaussian models produced a bias $b>1$ on scales $1-10
h^{-1}Mpc$. \footnote{Here and throughout this article, $h\equiv H_0$/(100
km/s/Mpc).}

This is particularly intriguing as earlier work (\cite{RobAlb}) has shown,
again in the context of cosmic string theories, that
dark matter pancakes, which arise in the non-linear evolution of 
Gaussian perturbations, effectively mimic the sheet-like overdensity 
of the wake. Hence, observing the density field alone will not lead to a
detection of primordial non-Gaussianity, even when tailor-made
statistics are being used.
This confusion
could be overcome if the interplay between gas dynamics and dark matter 
produced baryon bias as a remnant 
of early non-Gaussianity which survives until today.
Another exciting possibility is that non-Gaussianity may lead to a
natural explanation of the observed high baryon fraction in the centre
of rich clusters (\cite{comabias}; \cite{abellbias})
as conjectured by \cite{Sornborger}.

However, \cite{Sornborger} simulate sheet--like wakes
in a perfectly homogeneous background. Combined with the perfect
planar symmetry of their wake model, this induces translational 
symmetry in the plane parallel to the wake and mirror symmetry about
the wake plane, 
an idealisation which limits the phase space available to
the system.
Here, I consider a ``realistic'' scenario of wake formation in a
gas/HDM--mixture at early times ($\eta^* = 6
\eta_{eq}$).
To allow the system to explore the
full 3-D phase space in its evolution 
I use the full response of the 
linearised Einstein equations 
to a line source to compute the initial wake perturbation and model
the effect of early-time wakes in the simulation volume by the
addition of a Gaussian field background.
Also, instead of allowing continuous inflow of new material into the
simulation volume I impose periodic boundary conditions on the cube as
a crude model of the effects of compensation. Another effect I include
is cooling due to bremsstrahlung from hot gas.

The plan of this paper is as follows: in the following
section I discuss some physical background 
and the assumptions and methods used in 
my simulations. In the third section I present the simulation results
and the final section contains my conclusions.
\newpage
\section{Methods}\label{methods}
\subsection{Sheet-like structure}
\begin{figure}
\plotone{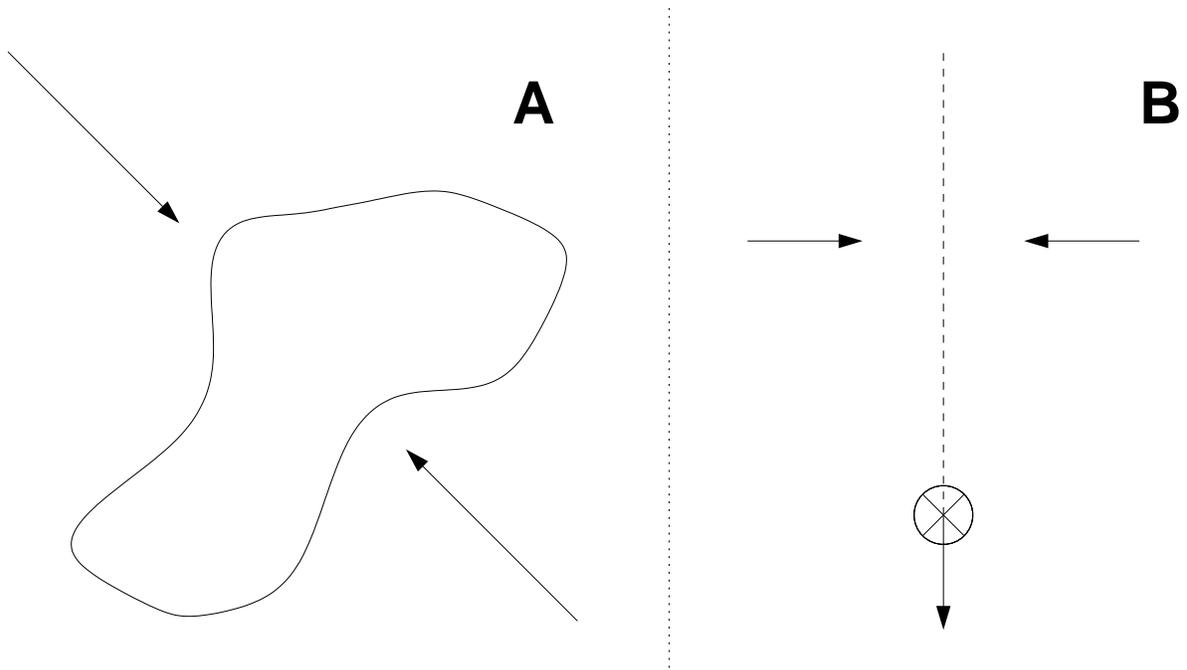} 
\figcaption[sheet.eps]{Two mechanisms for sheet formation. A) Non-linearly
growing density
perturbations form Zel'dovich pancakes. B) A string, marked by
$\otimes$ moving in the indicated direction gathers up material in its
wake (dashed).
\label{sheet}
}
\end{figure}

The reason why different physics may be expected in
cosmological sheets formed in the wake of a cosmic string and those
that are generated through gravitational instability of Gaussian
fluctuations is illustrated in Figure \ref{sheet}. It was shown by 
Zel'dovich, that most initially
overdense regions in a Gaussian scenario have a preferred direction of
collapse associated to them, which causes them to evolve into
cosmological pancakes. The characteristic scale 
of these pancakes and their overdensity are given by the shape and
amplitude of the
powerspectrum of initial fluctuations.

In the case of cosmic string wakes, the physical characteristics of
the wake are determined by a different 
dynamical process. Imagine a homogeneous
background density and a straight string moving through it. Through
the effect of the conical deficit of space-time around the string the
particles behind the string feel a velocity kick in the direction
perpendicular to and towards the plane swept out by the string.
In linear theory, the overdensity $\de$ which is thus created, 
can be written as the response
to the positive trace of the string stress energy tensor
$\Th_+\equiv \Th_{00}+\Th_{ii}$,
\beq
 \twid{\de}(\v{k})=4\pi \frac{(1+z_{eq})}{(1+z_{hydro})}
\int_{\eta_{i}}^{\eta_{f}}\twid{T}(k;\eta') 
\twid{\Th}_+(\v{k},\eta')d\eta'
\label{source}
\eeq
Here the source acts between conformal times ${\eta_{i}}$ and
${\eta_{f}}$, $\twid{T}(k;\eta)$ is the HDM-transfer function, and 
$z_{eq}=\frac{1}{1+a_{eq}}, a_{eq}=4.17\times 10^{-5} h^{-2}$. Fourier
transforming then yields the real-space perturbation $\de(\v{x})$.

The parameters which govern the shape and strength of 
the wake perturbation are the string mass
per unit length, the curvature scale of the string and 
its bulk velocity. These enter
equation \ref{source} through $\twid{\Th}_+$.
Hence, the physical processes which 
form cosmic string wakes are
very different from the Zel'dovich collapse of pancakes described above.

\subsection{Initial conditions}

The way I model the initial fluctuations derives from Robinson \&
Albrecht (1996). 
The density field consists of a non-Gaussian ``wake'' part and a Gaussian
part. To understand the appearance of the Gaussian part, consider the
following reasoning: a simple scaling argument shows that in a
co-moving frame and in a matter dominated era the 
string network appears to dilate and stretch with physical time as
$t^{1\over3}$ (more technically,
the curvature scale and the inter-string distance increase). This means
that at early times there is a dense tangle of strings inside the
box. The
integrated perturbations seeded by this tangle
are modeled as a Gaussian fluctuation
background with power-spectrum given by Albrecht \& Stebbins (1992b). 
Since many strings
source this perturbation, the use of a Gaussian model is suggested by
the central limit theorem.
In this work I focus on the I-model of cosmic strings, which broadly
agrees with recent determinations of galaxy power spectra (e.g.\
Albrecht et al. 1997a; Pen et al. 1997a). For this
purpose I chose the parameters shown in Table \ref{simpars}.

A single string 
seeds the non-Gaussian ``wake'' part at the time $\eta^* = 6
\eta_{eq}$. Why choose this particular time? This is just the
time when the string has the best chance of 
producing a wake which stands out against the Gaussian background in an
HDM-scenario in linear theory\footnote{As has been noted in
Robinson \& Albrecht (1996), this is true
for all three of the very different models of cosmic string which are
considered by Albrecht \& Stebbins (1992b).}. 
If I prefer a wake which is straight on the
scale of the box, picking this time determines the boxsize used
to be the
curvature scale of the string network at $\eta^*$. This evaluates to
$20 h^{-1}$ Mpc in the I-model.

\begin{figure}
\plotone{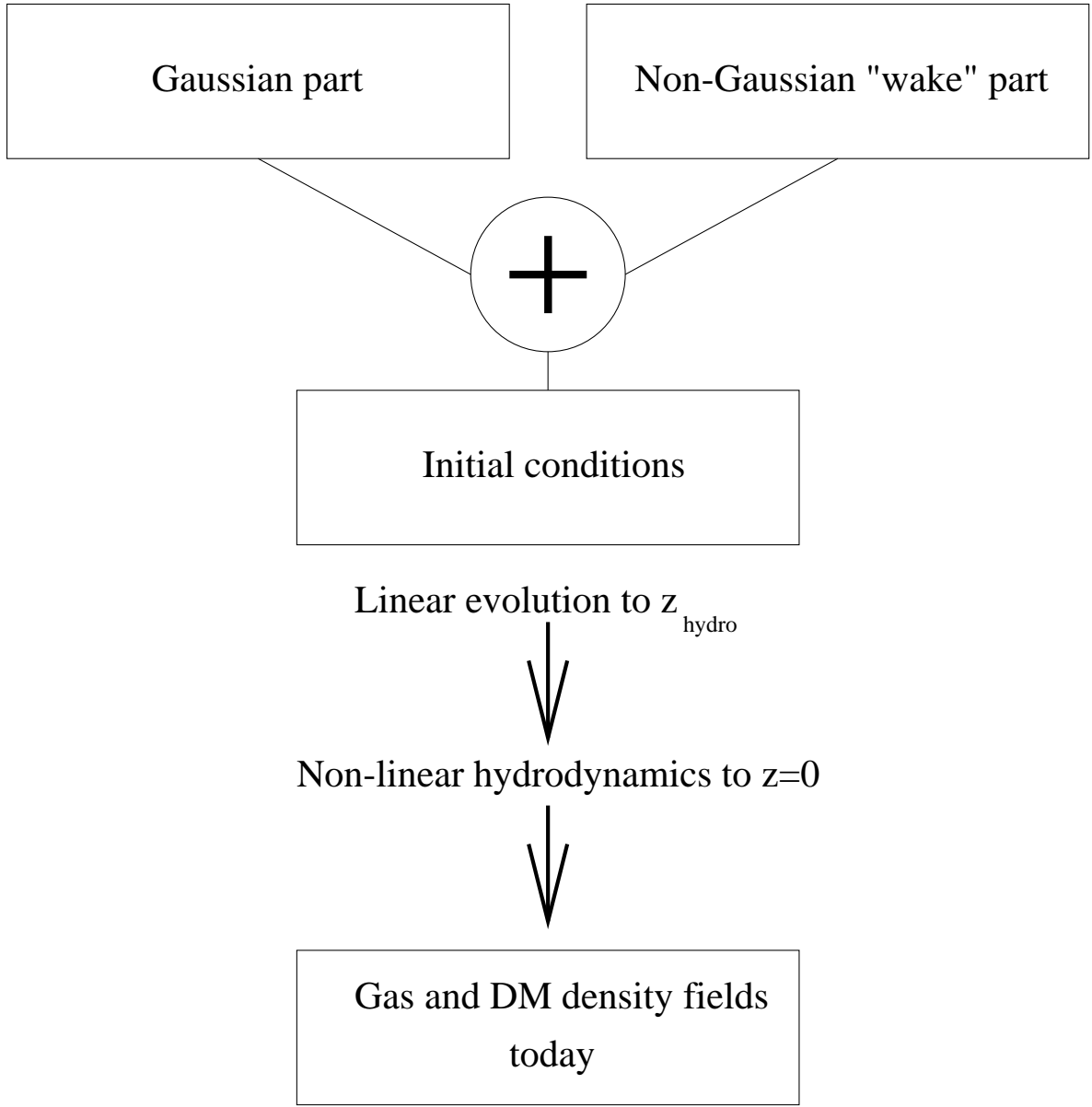} 
\figcaption[plan.eps]{A flow chart showing the overall plan of my simulations.
\label{plan}
}
\end{figure}

Seeing as the perturbations equations are still linear at this stage, I can
obtain the full initial conditions by simply adding together the
Gaussian and non-Gaussian parts.
Note that $\sigma_{wake (1 Mpc)}$, the fluctuation in a $1 h^{-1}$ Mpc
ball centred on the wake, is only 20\%  larger than 
the fluctuation in the Gaussian
field in a ball of the
same size. The Gaussian background is therefore expected to have a
non-negligible effect.
The simulation itself proceeds in two steps: First the
initial conditions are evolved to a redshift $z_{hydro}$ in linear
theory. Then they are fed into the hydro-code and evolved using the
full non-linear evolution equations until today $(z=0)$ (\cf Figure
\ref{plan}).

Treating much of the history of the wake in linear theory can be
justified by numerical and analytical studies of the planar symmetric,
1-D case (\cite{sorn2}) showing that HDM wakes formed
at conformal time $\eta_{eq}$ go 
non-linear only at a redshift $z_{nl}\simeq 30$.

This sets up the formalism for calculating the dark matter
perturbations in linear theory. So far nothing has been said about the
gas because a general formalism does not exist for calculating the
perturbations in the baryonic fluid.
Hence, at this stage one has to decide what initial conditions to
assign to the gas. I give it
the same initial perturbations as the dark matter fluid and justify
this choice by observing
that in Boltzmann calculations of the baryon overdensity one observes
the baryons to rapidly flow into the potential wells created by
the dark matter overdensities for a wide range of cosmological
parameters (\cite{MaBert}). For the value of
$z_{hydro}$ used in  my simulations I expect the baryonic density
field to follow the dark matter very closely. I have
tested this assumption by starting the hydrocode soon after decoupling
($z_{hydro}=1000$) and perturbing only the dark matter and leaving the
gas completely homogeneous.
I find that the results concerning the baryon bias in the centre of
the wake, baryon fractions in
dense peaks and wake fragmentation patterns at
$z=0$ remain largely unchanged under
this radical change in initial conditions.

\subsection{Non-linear evolution}

For the non-linear evolution of the overdensities in both the baryonic
and the dark matter fluids, I use the Adaptive Moving Mesh Algorithm 
developed by Pen (1995, 1997). This code has the following features:
\bi
\item Good shock resolution
\item High speed of execution
\item Good resolution of high density areas
\item Grid based TVD approach ensures optimal control of artificial 
viscosity
\ei

The Moving Mesh Hydrodynamic (MMH) method produces output on a
irregular, curvilinear 
grid. This means standard statistics such as the powerspectrum are not
easily calculated for these density fields. Fortunately, for this
application, I am more interested in understanding the dynamics of
dark matter and baryons in a
cosmic string wake and the processes leading to its
fragmentation. This information is not readily extractable from two-point
statistics. Instead I will look at various cuts through the data sets.
To use the particular strength of the MMH method, I am going to
focus on high density regions. 

While in actual fact the N-body solver which computes the dark matter
dynamics treats the dark matter as cold, 
a light massive
neutrino of mass $m_\nu=93$eV has
free-streamed a distance of 9 kpc since 
$z=100$. In the highest density areas the smallest mesh elements have
size 15.6 kpc, so the dark matter will be slightly too
compressed at the small scale resolution limit. This is not expected
to affect my results, since this effect acts on scales 
small compared with the width of the
wake.

Cooling due to bremsstrahlung is included in this code by computing
the amount of energy radiated away from each mesh cell in the
simulation volume. The free--free luminosity of the cell is calculated in the
standard way (\eg chapter 24 in \cite{peebles} and references therein)
but for the inclusion of a helium
abundance of $Y_{He}=0.24$ and the Gaunt factor $g=1.2$.

\section{Results and Discussion}
\label{results}

The code was run on COSMOS, the UK National Cosmology
Supercomputer. This is a Silicon Graphics Orgin 2000 with 32 R10000 
processors and a shared memory architecture \footnote{
For technical information about COSMOS refer to 
{\em http://www.damtp.cam.ac.uk/cosmos/home.html}}. I
simulated wakes in $\Omega_b=0.05$ and $\Omega_b=0.1$ universes and
completed 3 production runs each on a mesh with $64^3$ cells. 

\subsection{Checks}

Several test runs were performed to check that the
results are only weakly 
dependent on parameter choices and compiler switches for 
the hydro-code, like the compression limit on mesh distortions.
To avoid overcompression of the baryons,
I chose to use the code without enforcing exact energy conservation.
Also, I ascertained that runs
with the same random seeds for the Gaussian background fluctuations 
but without wakes showed no significant baryon bias. Tests of code
performance itself are published elsewhere (\cite{penbody}, 1997).

\subsection{Simulation results}

\begin{figure}
\epsscale{0.8}
\plotone{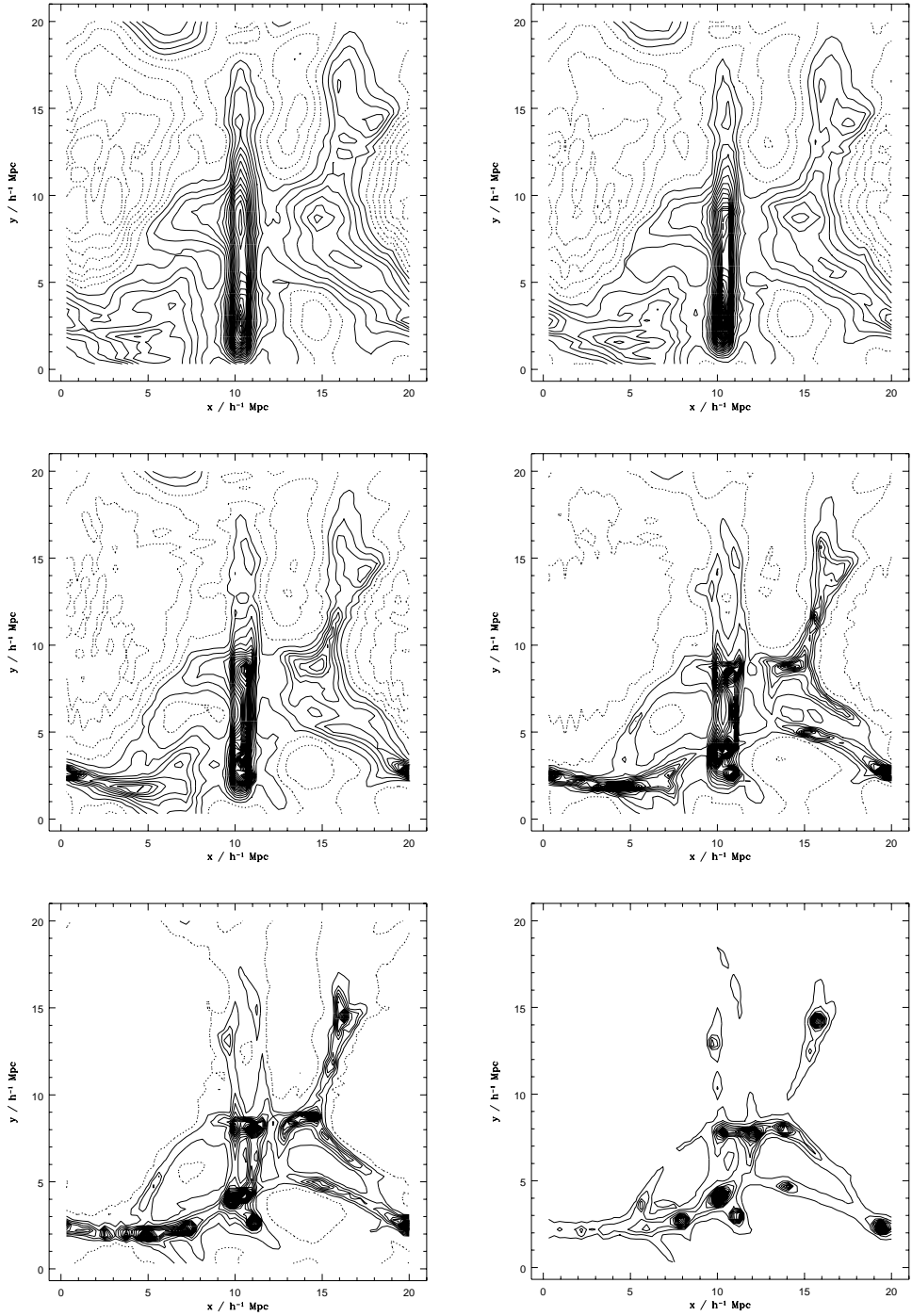}
\figcaption[gasZ2.eps,gasZ3.eps,gasZ4.eps,gasZ5.eps,gasZ6.eps,gasZ0.eps]
{The averaged CIC-smoothed density in the
gas. Snapshots are
taken at $z=8,4,2,1,0.5$ and 0, from top left to bottom right. The
wake can be clearly seen in the centre of the box at early times 
and later structure emerges
from the Gaussian perturbation background. The 28 contours are
equally spaced in smoothed averaged density from 1.47 to 36.64.
\label{gasevol.fig}
}
\end{figure}

\begin{figure}
\epsscale{0.8}
\plotone{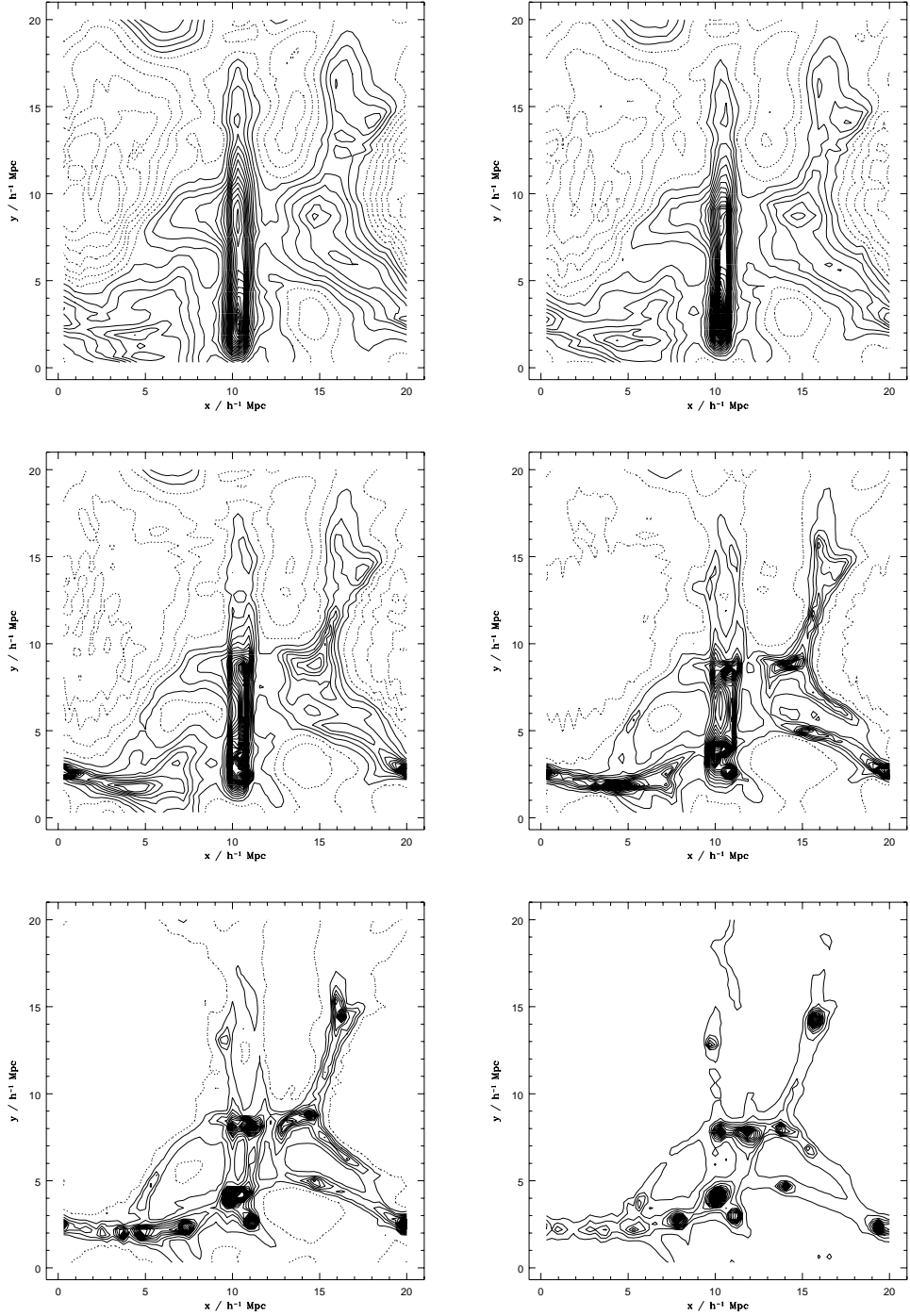}
\figcaption[darkZ2.eps,darkZ3.eps,darkZ4.eps,darkZ5.eps,darkZ6.eps,darkZ0.eps]
{As Figure \protect\ref{gasevol.fig} but for the dark matter. Here,
the 28 contours range from 1.31 to 32.30.
\label{darkevol.fig}
}
\end{figure}

\begin{figure}
\epsscale{0.8}
\plotone{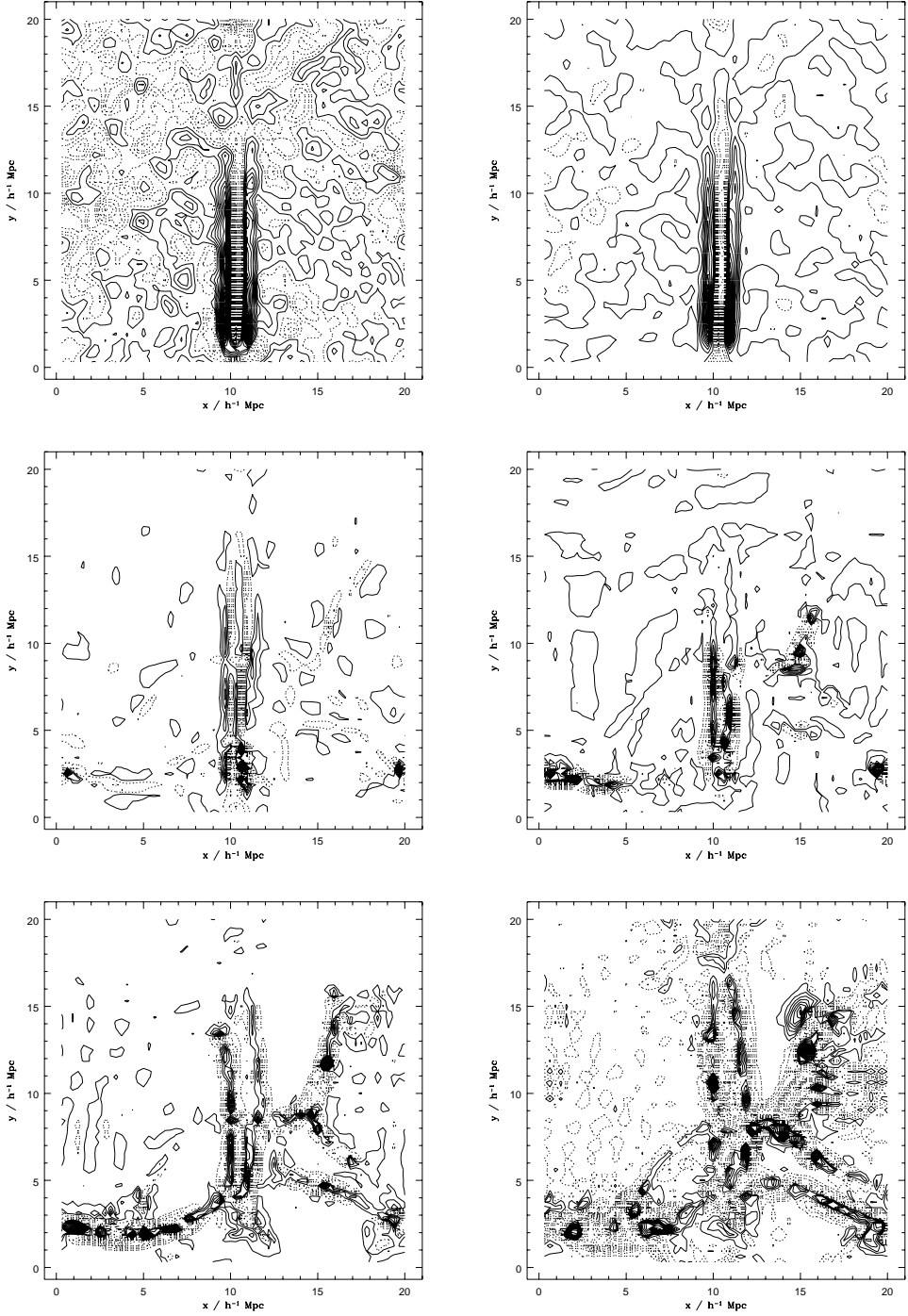}
\figcaption[biasZ2.eps,biasZ3.eps,biasZ4.eps,biasZ5.eps,biasZ6.eps,biasZ0.eps]
{As Figure \protect\ref{gasevol.fig} but for the fraction of
gas to dark matter densities. At late times the wake becomes warped
due to neighbouring fluctuations, which causes the apparent doubling
of the wake plane (\cf Figure \protect\ref{biasslice.fig} for a
cut through the $z$=0 volume). The 28 contours range from 0.50 to 1.97.
\label{biasevol.fig}
}
\end{figure}

As expected, adding a Gaussian background to the theory makes the
results considerably richer than in the symmetric case.
I find a complex interplay of gravitational and hydrodynamic forces
in string wakes. 
To give a general idea about the output from such simulations, Figures
\ref{gasevol.fig}, \ref{darkevol.fig} and \ref{biasevol.fig} show
the simulation volume for one realisation, 
averaged over one direction along the wake and
looking edge-on at the wake. Snapshots of the simulation volume
are taken at $z$=8, 4, 2, 1, 0.5 and 0.
As the code output is non-uniformly sampled, it has to be smoothed
appropriately for display purposes.
The density field is computed at the mesh cell centres. These are 
rebinned onto a cubic lattice with weighting factors determined by the
Cloud-in-Cell scheme (CIC). This prescription is going to work well in
regions of high density where the density of mesh cells is enhanced by
a factor of up to several 1000 compared to the original cell density, whereas
it is going to lead to spurious peaks close to the mesh cell centres 
in highly underdense regions.

The approximate symmetry along the wake can be exploited 
by averaging through the box 
to effect further smoothing and to obtain 2-D
fields for easy visualisation.
This has the advantage
over slices through the volume that all high density areas will be
seen, but may wash out features on the wake. 1-D profiles of the wake
are obtained by another average in the 
direction of cosmic string propagation.

I measure an average of 70\% maximum baryon excess at $z$=0, 
after CIC-binning, in the centre
of the wake. This is about half
of what
\cite{Sornborger} find but there are significant differences
between their simulation setup and mine. Adding a  Gaussian
background would be expected to reduce the coherence in the dark
matter oscillations about the wake plane and so would the density
gradient along the wake. Also, my crude model of compensation
means that the inflow of new material into
the wake is going to subside at late times. Last but not least, the
string bulk velocity is smaller in my simulations. All these effects are
expected to reduce the baryon bias. Maybe it should be stressed that
it is surprising that any effect survives at all, which
underlines the robustness of this prediction. 

Figure \ref{gasevol.fig} shows the evolution of the 
gas perturbations for one realisation. The wake
is clearly visible in the first panel. The field is oriented such that
the string entered at the bottom centre of the plot and propagated
through until reaching the top centre. The density gradient along the
direction of wake propagation is due to the fact that matter has had
more time to fall in towards the centre where the string entered the
box. 
Subsequent gravitational
evolution leads to the fragmentation and collapse of the sheet. It can
be clearly seen how the Gaussian background induces the disruption of
the initial approximate symmetry. While it is still true that the
majority of high density peaks form on the wake, the sheetiness of the wake is
masked by other  sheet-like features which form in the final
frames. 
The dark matter distribution in Figure 
\ref{darkevol.fig} looks visually similar at all times.

\begin{figure}
\epsscale{0.63}
\plotone{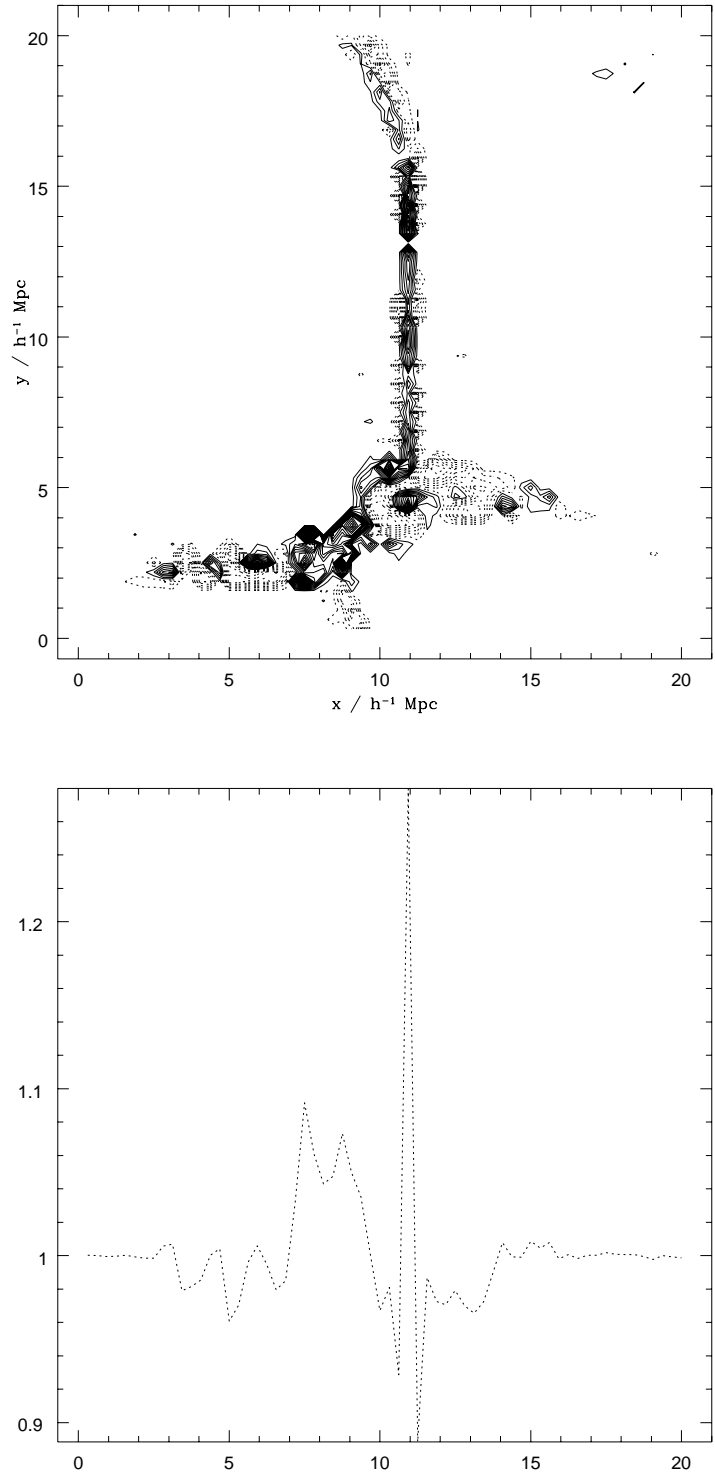}
\figcaption[biasslice2DZ0.eps,biassliceZ0.eps]
{The top panel shows a slice which was taken through 
the simulation volume in
the last panel of Figure \protect\ref{gasevol.fig}. The average of the
top slice in the y direction is shown in the bottom panel.
\label{biasslice.fig}
}
\end{figure}

The fraction of baryonic mass to dark mass per smoothed 
cell is shown in Figure 
\ref{biasevol.fig}. Encouragingly, the results broadly agree with the
calculations by \cite{Sornborger}.
The apparent doubling of the
wake structure after $z=2$ should be attributed to the distortion of the
wake plane due to the presence of neighbouring perturbations (remember
that these plots are averages through the volume). Looking at the cut
through the simulation volume at $z=0$ in Figure \ref{biasslice.fig} reveals
the baryon enhancement in the centre of the wake. 
\footnote{
Fluctuations in baryon bias away from the wake are less pronounced and
originate from the caustics that form when pancakes collapse.}
The wake thickness can be read off to be about $1.5 h^{-1}$ Mpc, which
is in very good agreement with earlier determinations.

\subsection{Baryon fraction in density peaks}

Another matter of interest is whether the baryon enhancement I observe
in the wake plane carries through to bound objects. 
The MMH method is particularly suited for addressing questions about high
density regions, since these are resolved very finely indeed. I 
select peaks by searching for local maxima and then
eliminating all those which are within $2h^{-1}$ Mpc of each
other to avoid double counting.
I then calculate the excess in the baryon fraction in a sphere of
$1h^{-1}$ Mpc radius around each peak over the
cosmological value. The population of peaks is separated in those
which are within $5h^{-1}$ Mpc of the wake plane and those further away. 
The results are summarised in Table \ref{baryfrac}. I find no
significant baryon excess in the highest density peaks in the wake
region, independent of the value of $\Omega_b$.

How can this be explained? I suggest that the spherical
collapse of high density peaks erases the initial baryon bias
caused by  the peculiarities of wake formation.
The changing geometry of the density field drives the following mechanism:

At first the baryons are tightly bound in the
Newtonian gravitational potential of the planar wake, 
$\psi_{plane}(x) \propto x$, 
where x is the distance from the wake plane. Assuming that at late
times the gas
is isothermal and in equilibrium in this potential simplifies the
hydrodynamic equations, and there is a simple analytic 
solution for the 1-D density profile,
\beq
\rho(x)=\frac{M}{2 L_{sheet}}{\rm sech}^2\left(\frac{x}{L_{sheet}}\right)
\label{profile}
\eeq
where $M$ is the mass per unit area of the sheet and $2L_{sheet}$ is
the its thickness (for a detailed derivation refer to 
Appendix \ref{planar}). This density falls off exponentially
away from the wake. As long as the 1-D symmetry is a good
approximation, the dark matter remains in the oscillatory mode which
was found by Sornborger (1997) and is 
thus prevented from reaching the
isothermal configuration. 

When a density peak starts
to form, the near-planar symmetry is supplanted by a near-spherical
symmetry in the vicinity of the peak. This gives rise to a point
charge potential $\psi_{point}(r) \propto \frac{1}{r}$ and the
isothermal configuration is now less tightly bound, as the asymptote of
the density decays as $r^{-2}$ (e.g.\ \cite{peebles}).
At the same time, the symmetry change allows the
dark matter fluid to escape the oscillatory mode and settle down to
the isothermal equilibrium distribution as well. The resulting inflow of dark
matter erases the
baryon excess in the peak.

\begin{figure}
\epsscale{0.63}
\plotone{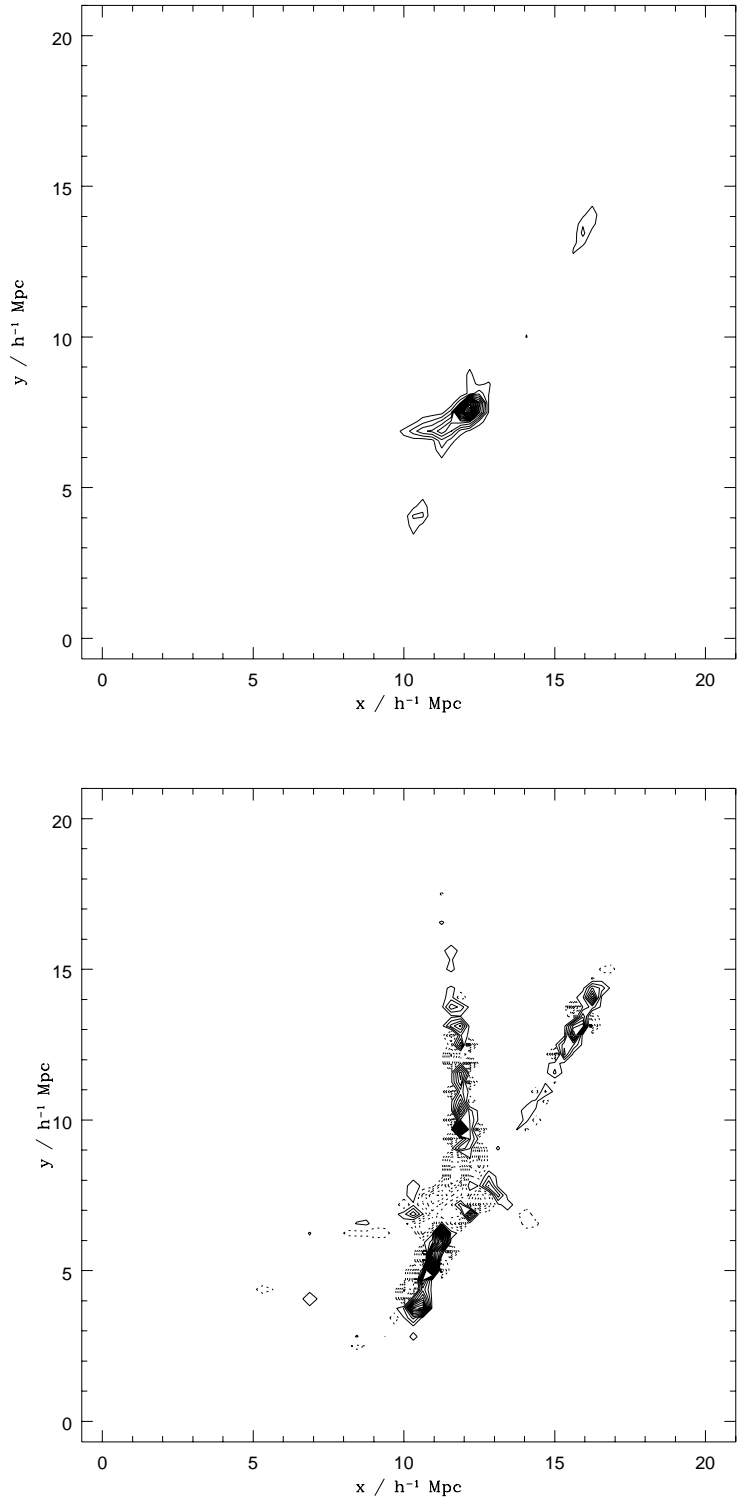}
\figcaption[peakbary.eps,peakbias.eps]
{A density peak on the wake. Shown is a slice through the
simulation volume, with the gas density in the top panel and the
fraction of baryon to dark matter density in the bottom panel. The
density peak punched a hole in the sheet of baryon excess.
\label{peak.fig}
}
\end{figure}

This explanation is corroborated when one examines the
simulation volume itself. Often baryon excess is seen {\em
surrounding} bound
objects at the end of the simulation, indicating that dark matter
flowed from the outside into the centre of the object. An example can
be seen in Figure \ref{peak.fig}: a slice through
the simulation volume that contains the object at
coordinate positions (12,7.5) in the final frame of Figure
\ref{gasevol.fig} shows that it is close to the centre of the
wake (warped away from the centre of the box by the other
fluctuations) and punches a hole into the sheet of baryon bias.

\section{Conclusions and implications for large scale structure}
\label{conclusions}

I studied a realistic scenario of cosmic string wake formation with a
view to understanding the differences in gas and dark matter dynamics
which are caused by the peculiarities of wakes. For this purpose I used
a state-of-the-art hydro code at high resolution. I find a wake
thickness of $1.5 h^{-1}$ Mpc and an average
maximum of baryon excess of 70\% in the core of the wake at $z=$0.

This baryon enhancement
does not carry through to baryon fractions in bound
objects. I proposed a mechanism for the erasure of baryon excess in
bound objects, which relies on geometrical factors related to the
symmetry change in the fragmentation of a cosmic string wake. This
model leads to improved understanding of object formation in these
non-standard scenarios. This could also be seen as strengthening the
case against $\Omega_m=1$ universes or for low Hubble constants, 
if one invokes the nucleosynthesis
bound on $\Omega_b$.

What do these simulations have to say about the relationship between
galaxy and dark matter powerspectra in cosmic string models? 
Definite statements would have to
await cosmic string network simulations which are coupled to a
hydrodynamics code and probe larger scales of up to 100 Mpc where the
discrepancy between the predicted cosmic string dark matter 
powerspectrum and the
observed galaxy powerspectrum is most acute. However, 
eye-ball results speak for a  close correspondence
of peaks in the gas and dark matter fields in the neighbourhood
of  a single wake. This discourages the thought that 
cosmic strings serve to differentiate the clustering properties of 
galaxies and dark matter. 

The last hope for favouring galaxy bias in large scale structure
generated by cosmic strings may be that the baryon--rich
environment of the inner wake region boosts galaxy formation
compared to other dense regions. However, the way in which baryon
excess is erased in density peaks 
serves as a reminder that two
sheets which are rich in dark matter sandwich the inner wake region. 
Thus, an abundance of dark matter is never far away and 
density peaks appear to play a prominent
r\^ole in mixing the baryonic and dark matter fluids at late times.

Even if the thin sheet of baryon bias could lead to an enhancement in galaxy
formation, it would be difficult to see how this would affect the
galaxy powerspectrum at $100 h^{-1}$ Mpc scales. Two scales are
important for the wake: the thickness across it and the size along
it. The prediction of the I-model for the size of straight sections of
dominant wakes is
$20 h^{-1}$ Mpc and the wake thickness is $1.5 h^{-1}$ Mpc. The
powerspectrum  even of an extreme  density field 
made out of cells with thin walls
of this characteristic size covered with ``galaxy wallpaper'',
would cut off
above $20 h^{-1}$ Mpc, failing to contribute to larger scale
power\footnote{However, it should be remembered that the size of 
dominant wakes
is model dependent. While detailed simulations are still
outstanding, other models of cosmic string may
predict somewhat larger values. 
Still, even a size of 60 Mpc must be considered extreme.}.

In summary, the cosmic string scenario fails to provide a natural
explanation for the high baryon fractions which are observed in
rich clusters. While this work is not directly relevant to the
$b_{100}$ problem posed by Albrecht et al. (1997a), I presented heuristic 
arguments why the peculiarities of wakes may be less important
to galaxy formation and biasing of the powerspectrum on large
scales than hitherto conjectured.

\acknowledgements
I am indebted to U.-L. Pen for 
making his hydrodynamics code available
to me. I would also like to thank A. Albrecht, J. Bartlett, 
N. Gnedin, U.-L. Pen,
A. Sornborger, and A. Stebbins for stimulating discussions and the
referee, R. Brandenberger, for fruitful comments.
I acknowledge the Knowles Studentship of the University of London and
support by the UK High Performance Computing Consortium, who granted me
access to COSMOS, the UK National Cosmology Supercomputer.

\appendix
\section{The Isothermal Sheet}
\label{planar}

The 1-D Boltzmann equation for an ideal gas in terms of the phase
space density $f(x,v)$ is
\beq
\dby{t}f(x,v)=0 .
\label{boltz}
\eeq
Let the density  $\rho(x)=\int f(x,v) dv$ and the average
streaming velocity $\avg{v}=\frac{1}{\rho}\int v f(x,v) dv$.
The zeroth and first moments of \eq{boltz} are (see \eg \cite{bt})
\beqa
\pdby{t}\rho +\pdby{x}(\rho \avg{v}) & = & 0 
\label{continuity}\\
\rho \pdby{t}\avg{v} + \rho \avg{v} \pdby{x} \avg{v} & = & 
-4 \pi G \rho \int_0^x \rho(r) dr -
\pdby{x}(\rho \avg{v^2} - \rho \avg{v}^2)
\label{euler}
\eeqa
Using the ideal gas equation of state identifies $\rho \avg{v^2}$ as
the gas pressure and $\avg{v^2}=\frac{k_B T}{m}$, where $m$ is the mass
of a  gas particle, $T$ is the gas temperature and $k_B$ is
Boltzmann's constant.

As the simulations show, close to the centre of the wake the baryons
sit in a pressure supported peak until spherical collapse sets in. 
To approximate the 1-D peak profile close to the centre, I assume 
that the baryons have interacted sufficiently to be in thermal
equilibrium and that streaming velocities can be neglected
compared to thermal effects. Hence $\avg{v}\simeq 0$ and using
the isothermal assumption $\pdby{x}\avg{v^2}=0$ gives
\beq
\avg{v^2}\pdby{x} \rho = - 4 \pi G \rho \int_0^x\rho(r) dr
\eeq
Differentiation yields
\beq
\pddby{x}\ln \rho= - \frac{4 \pi G}{\avg{v^2}} \rho
\eeq
which, together with the condition that the maximum density be at the origin,
has the solution 
\beq
\rho(x)=C \frac{k_B T}{8\pi G m} {\rm sech}^2\left(\frac{\sqrt{C} x}{2}\right)
\eeq
To fix the remaining constant of integration C, let the mass per unit
area in the
sheet be 
$$M\equiv\int_{-\infty}^{+\infty}\rho(r) dr=\frac{k_B T
\sqrt{C}}{2 \pi G m}$$
 Then
\beq
\rho(x)=\frac{M}{2 L_{sheet}}{\rm sech}^2\left(\frac{x}{L_{sheet}}\right)
\eeq
where $2 L_{sheet}=\frac{2 k_B T}{\pi G m M}$ is the thickness of the
sheet.

These formulae reproduce the scaling of $M$ with
$L_{sheet}$ I observe in my simulations. Typically, for
$\Omega_b=0.05$ and at $z=0$, a fifth of the total mass of the box,
that is $8.6\times 10^{47}$ g, is within the inner wake region.
This gives $M=2.4\times 10^{-4}\frac{\rm g}{\rm cm^2}$. 
$T\simeq 1.4\times 10^6 K$ and hence $2 L_{sheet}\simeq 4.4\times
10^{24}$cm$\simeq 1.4$ Mpc, very close to the observed thickness of the
wake.

\newpage

\begin{table}[f]
   \begin{center}
        \begin{tabular}{|c|c|c|}
                \tableline
                Boxsize &  $L$ & $20 h^{-1}Mpc$ \\

                Redshift  & $z_{hydro}$ & 100 \\
                 
		Wake formation & $\eta_*$ & $6\eta_{eq}$\\
                 
		Mass/unit length & $\mu_6$ & $1.1$\\
                 
		String bulk velocity & $\beta$ & $.3c$ \\
                
		Hubble parameter & $ h_0$ & 1 \\
                
		Density parameter & $\O$  & 1\\
		
		Baryon density & $\O_b$ & 0.05 and 0.1\\
                \tableline 
		
        \end{tabular}
        \caption[Simulation parameters]
        {The parameters used in the hydrodynamics simulations. Three
        runs were done for each value of the baryon density.\label{simpars}}
\end{center}
\end{table}
\newpage

\begin{table}[f]
   \begin{center}
        \begin{tabular}{|l|c|c|}
                \tableline 
                
                Baryon excess... & $\Omega_b=0.05$ & $\Omega_b=0.1$ \\
                \tableline 
                \tableline 
                 
                ... within $5h^{-1}$ Mpc of wake plane & $1.03\pm 0.09$
        	& $1.09\pm 0.156$\\
                 
		... away from wake & $ 1.08\pm 0.16$ & $1.035\pm.061$\\
                \tableline 
        \end{tabular}
        \caption[Baryon fractions in high density regions]
        {This table shows the excess of baryon fractions in density
                peaks over the cosmological value with 1-$\sigma$ standard
        deviations. \label{baryfrac} }
\end{center}
\end{table}
\newpage


\newpage

\newpage

\newpage

\newpage

\newpage

\newpage

\newpage


\begin{thebibliography}{99}

\bibitem[Albrecht, Battye \& Robinson 1997a]{ABR1} 
Albrecht, A., Battye, R., \& Robinson, J.\ H.\ 1997,
Phys. Rev. Lett. 79, 4736

\bibitem[Albrecht, Battye \& Robinson 1997b]{ABR2} 
Albrecht, A., Battye, R., \& Robinson, J.\ H.\ 1997,
astro-ph/9711121, preprint

\bibitem[Albrecht \& Stebbins 1992a]{A&Sa}
Albrecht. A.\ \& Stebbins, A.\ 1992, Phys. Rev. Lett., 68, 2121 

\bibitem[Albrecht \& Stebbins 1992b]{A&Sb}
Albrecht. A.\ \& Stebbins, A.\ 1992, Phys. Rev. Lett., 69, 2615 

\bibitem[Binney \& Tremaine (1987)]{bt} 
Binney, J.\ \& Tremaine, S.\ 1987, Galactic Dynamics
(Princeton: Princeton University Press)

\bibitem[Jenkins et al. 1997 ]{virgo} 
Jenkins, A., et al. 1997, to appear 
in Dark and Visible Matter in Galaxies and Cosmological
Implications, ed.  M.\ Persic \&  P.\ Salucci, PASP conference Series. 

\bibitem[Kibble 1976] {kibble} 
Kibble, T.\ W.\ B.\ 1976, J. Phys., A9, 1387 

\bibitem[Kolb \& Turner (1990)]{KandT} 
Kolb, E.\ W.\ \& Turner, M.\ S.\ 1990, The Early
Universe (Reading: Addison-Wesley)

\bibitem[Loewenstein \& Mushotzky 1996]{abellbias} 
Loewenstein, M.\ \& Mushotzky, R.\ F.\ 1996, astro-ph/9608111 preprint 

\bibitem[Ma \& Bertschinger 1995]{MaBert} 
Ma, C.\ P.\ \& Bertschinger, E.\ 1995, Ap.J. 455, 7

\bibitem[M\"ah\"onen 1996]{maeh} M\"ah\"onen, P. 1996,
Ap.\ J.\ 459, L45

\bibitem[Peebles (1993)]{peebles}
Peebles, P.\ J.\ E.\ 1993,  Principles of physical cosmology
(Princeton: Princeton University Press)

\bibitem[Pen 1995]{penbody}  Pen, U.--L.\ 1995, Ap.J.S., 100, 269

\bibitem[Pen 1997]{pencode} Pen, U.--L.\ 1997, astro-ph/9704258,
Ap.J.S., submitted

\bibitem[Pen, Seljak \& Turok 1997]{PST1}  
Pen, U.--L., Seljak, U.\ \& Turok,N.\ 1997, Phys. Rev. Lett. 79, 1611

\bibitem[Perivolaropoulos, Brandenberger \& Stebbins 1990]{perivolaro}
Perivolaropoulos, L., Brandenberger, R.\ H.\ \& Stebbins, A.\ 1990,
Phys. Rev. D41, 1764

\bibitem[Robinson \& Albrecht 1996]{RobAlb} 
Robinson, J.\ H.\ \& Albrecht, A.\ 1996, M.N.R.A.S., submitted

\bibitem[Sornborger 1997]{sorn2} Sornborger, A.\ 1997, Phys.\ Rev.\ D56, 6139

\bibitem[Sornborger et al. (1997)]{Sornborger} 
Sornborger, A., et al.\ 1997, Ap.\ J.\ 482, 22

\bibitem[Stebbins et al. 1987]{stebbETAL} 
Stebbins, A.\ et al.\ 1987, Ap.\ J.\ 322, 1

\bibitem[Turok, Seljak \& Pen 1997]{PST2}   
Turok, N.\, Pen, U.--L.\ \& Seljak, U.\ 1997, astro-ph/9706250, preprint

\bibitem[Vachaspati 1986]{Vachaspati1}
 Vachaspati, T.\ 1986, Phys.\ Rev.\ Lett.\ 57, 1655

\bibitem[Vachaspati 1992]{Vachaspati} 
Vachaspati, T.\ 1992, Phys.\ Rev.\ D45, 3487

\bibitem[Vachaspati \& Vilenkin 1991]{VachVil}
Vachaspati, T.\ \& Vilenkin, A.\ 1991, Phys. Rev. Lett., 67, 1057

\bibitem[Vilenkin \& Shellard 1994]{V&S} 
Vilenkin, A.\ \&  Shellard, P.\ 1994, Cosmic Strings and
other Topological Defects
(Cambridge: Cambridge University Press)

\bibitem[White et al. 1993] {comabias} White, S.\, et al. 1993, Nature
366, 420

\end{thebibliography}
\end{document}